\documentclass{PoS}

\title{Two aspects of the Regge limit in QCD: Double Logs in Exclusive observables and Infrared Effects in Cross Sections}

\ShortTitle{Two aspects of the Regge limit in QCD: Double Logs in Exclusive observables and Infrared Effects in Cross Sections}

\author{{Grigorios Chachamis}$^1$, {Douglas A. Ross}$^2$, \speaker{Agust{\' \i}n Sabio Vera}$^1$
\thanks{DAR thanks the Leverhulme Trust for an Emeritus Fellowship.  GC and ASV acknowledge support from MICINN (FPA2016-78022-P). This work is supported by the Spanish Research Agency through the grant IFT Centro de Excelencia Severo Ochoa SEV-2016-0597.}
\\
$^1$ {\small Instituto de F{\' \i}sica Te{\' o}rica UAM/CSIC}
{\small \& Universidad Aut{\' o}noma de Madrid, Madrid, Spain.}\\
$^2$ School of Physics \& Astronomy, University of Southampton, Southampton SO17 1BJ, UK\\
       E-mail: \email{a.sabio.vera@gmail.com}}

\abstract{Two relevant points related to the application of the BFKL formalism to phenomenology are discussed. First, we have presented a set of observables characterizing multi-jet configurations event by event (average transverse momentum, average azimuthal angle, average ratio of jet rapidities) which can be used to find distinct signals of BFKL dynamics at the LHC. A numerical analysis has been shown using the Monte Carlo event generator {\tt BFKLex}, modified to include higher-order collinear corrections in addition to the transverse-momentum implementation of the NLO kernel. We require to have two tagged forward/backward jets in the final state. Second, the structure of the BFKL equation changes if infrared boundary conditions are imposed when considering the running of the coupling. The cut  in the complex angular momentum plane becomes an infinite series of Regge poles. Integrating along a contour off the real axis we find a strong dependence of the intercepts and collinear regions on the choice of the boundary conditions.  The mean transverse scales dominant in the gluon ladder increase. This could have interesting phenomenological consequences. }

\FullConference{ 25th International Workshop on Deep Inelastic Scattering and Related Topics \\
                 3-7 April 2017 \\
                 University of Birmingham, Birmingham, UK}

\begin{document}

\section{Double Logs in Exclusive observables}

\begin{figure}
\begin{center}
\includegraphics[height=7cm]{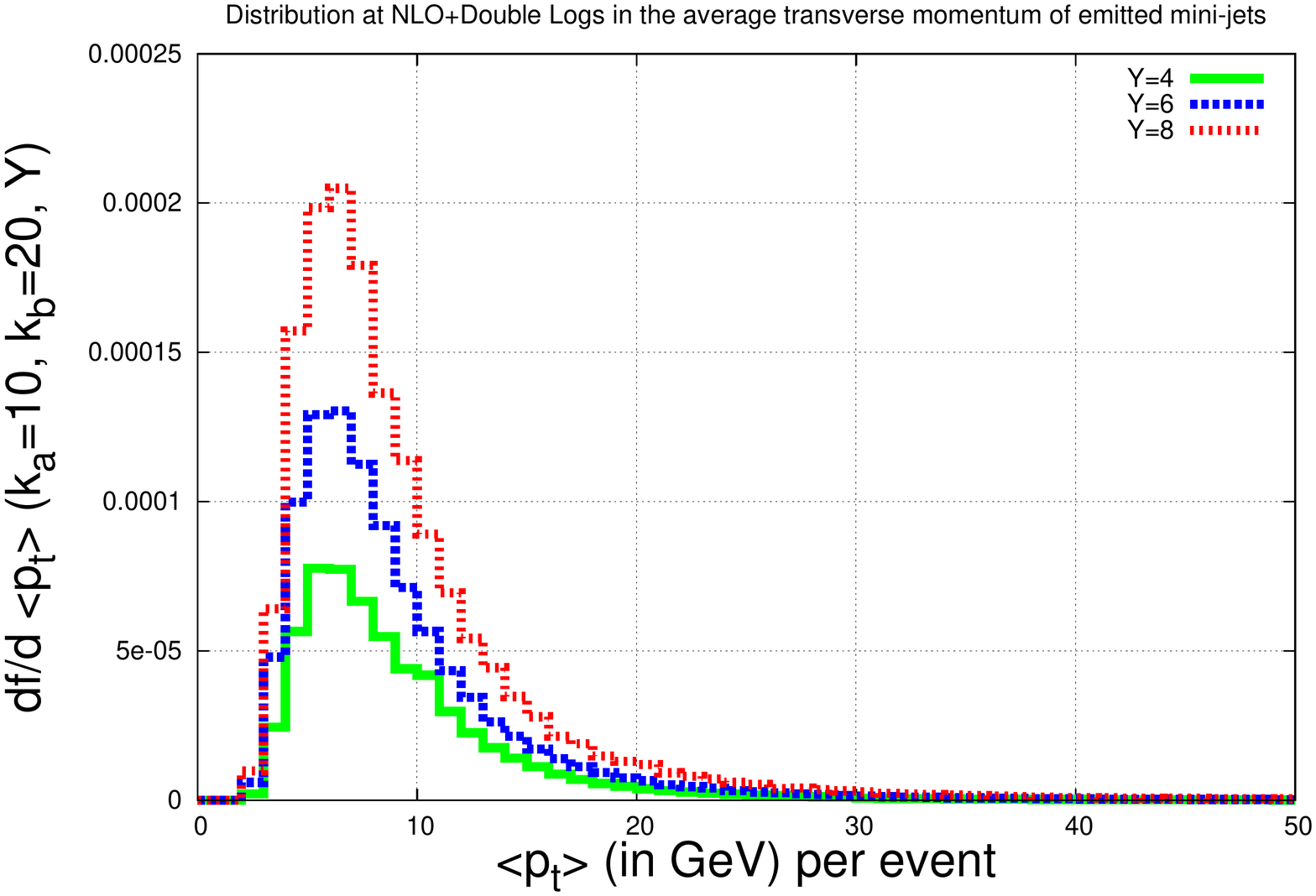}
\includegraphics[height=7cm]{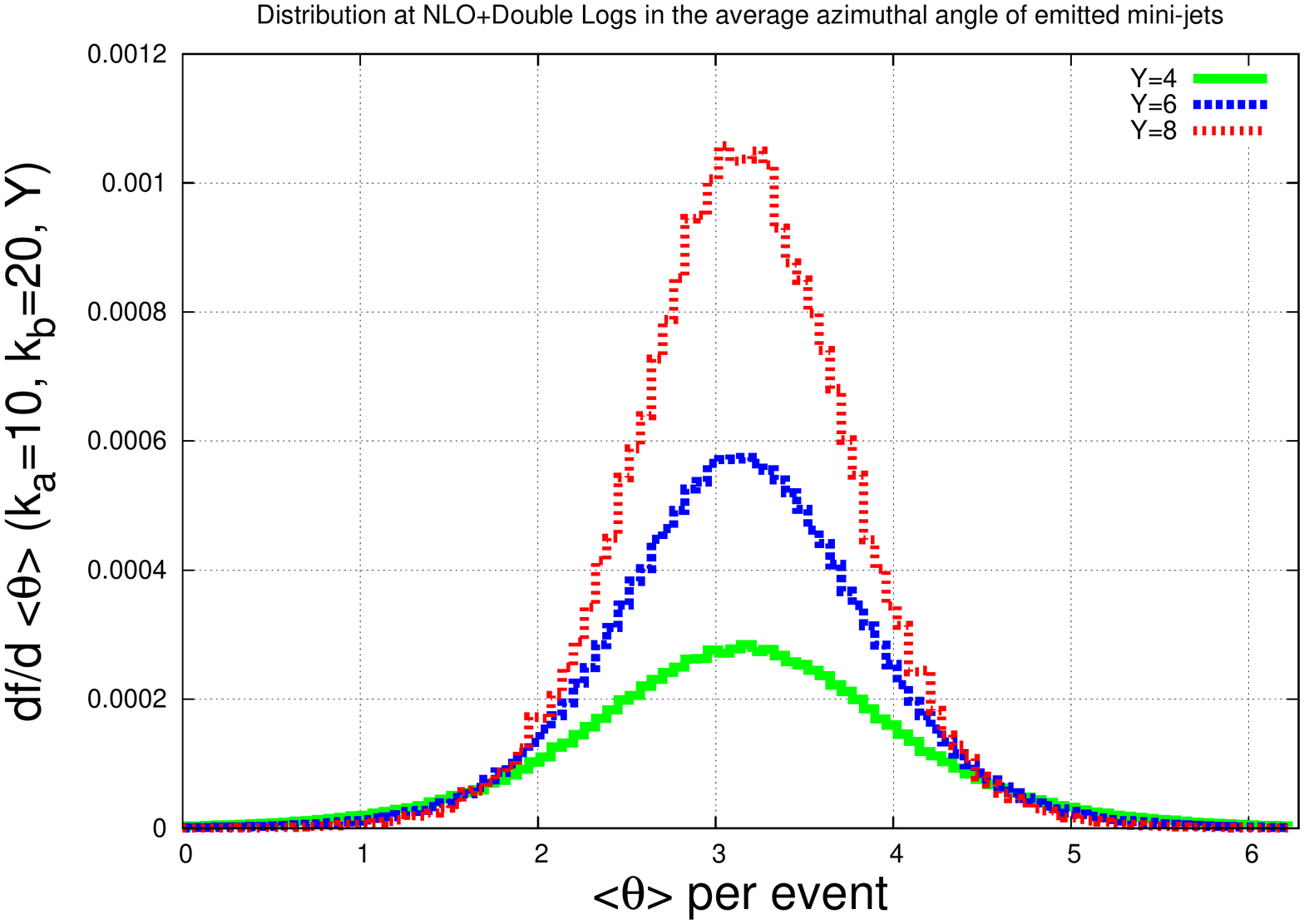}
\includegraphics[height=7cm]{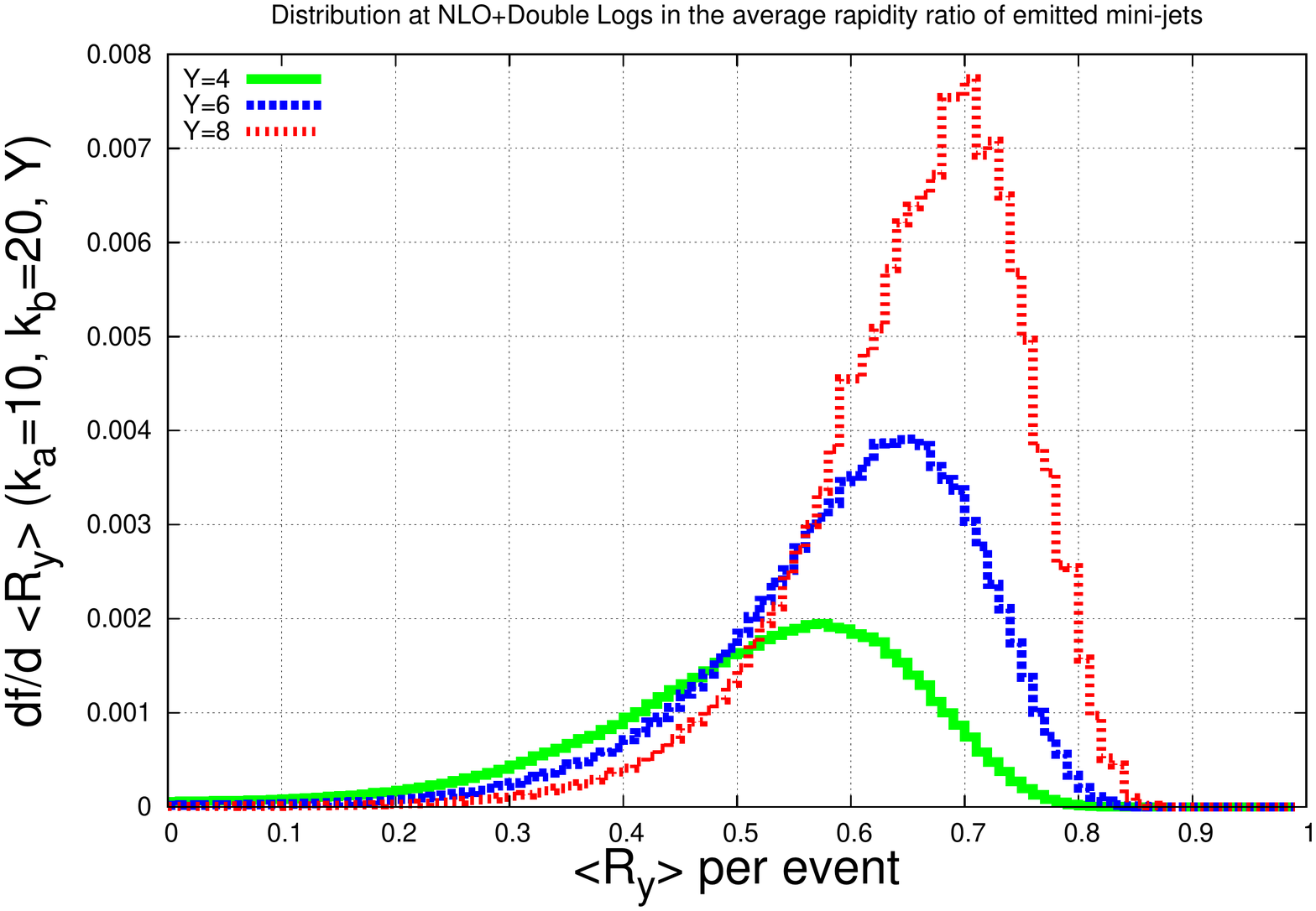}
\end{center}
\vspace{-.4cm}
\caption{High-multiplicity distributions generated with {\tt BFKLex}.}
\label{Plots}
\end{figure}

The BFKL~\cite{BFKL} approach allows for the study of high-multiplicity final-states even if the scattering energy is not large. In events where two forward/backward jets are tagged cross sections  are 
\begin{eqnarray}
\sigma (Q_1,Q_2,Y) = \int d^2 \vec{k}_A d^2 \vec{k}_B \, {\phi_A(Q_1,\vec{k}_a) \, 
\phi_B(Q_2,\vec{k}_b)} \, {f (\vec{k}_a,\vec{k}_b,Y)}.
\end{eqnarray}
$\phi_{A,B}$ are impact factors depending on scales, $Q_{1,2}$.  The gluon Green function $f$ depends on $\vec{k}_{a,b}$ and the center-of-mass energy in the scattering. At NLO we include running of the coupling and energy scale in the resummed logarithms
~\cite{Forshaw:2000hv}.
%~\cite{Forshaw:2000hv,Chachamis:2004ab,Forshaw:1999xm,Schmidt:1999mz}.
 $f$ can be written in the iterative form
 ~\cite{Schmidt:1996fg}
% ~\cite{Schmidt:1996fg,Andersen:2003an,Andersen:2003wy}
\begin{eqnarray}
f &=& e^{\omega \left(\vec{k}_A\right) Y}  \Bigg\{\delta^{(2)} \left(\vec{k}_A-\vec{k}_B\right) + \sum_{n=1}^\infty \prod_{i=1}^n \frac{\alpha_s N_c}{\pi}  \int d^2 \vec{k}_i  
\frac{\theta\left(k_i^2-\lambda^2\right)}{\pi k_i^2} \nonumber\\
&\times& \int_0^{y_{i-1}} \hspace{-.3cm}d y_i e^{\left(\omega \left(\vec{k}_A+\sum_{l=1}^i \vec{k}_l\right) -\omega \left(\vec{k}_A+\sum_{l=1}^{i-1} \vec{k}_l\right)\right) y_i} \delta^{(2)} \hspace{-.16cm}
\left(\vec{k}_A+ \sum_{l=1}^n \vec{k}_l - \vec{k}_B\right) \hspace{-.2cm}\Bigg\}, 
 \end{eqnarray}
where $\omega \left(\vec{q}\right)$ is the gluon Regge trajectory and $\lambda$ regularizes the infrared divergencies. The Monte Carlo event generator {\tt BFKLex} implements this iteration and can be used for collider phenomenology and more formal studies~\cite{Chachamis:2013rca}. 
%~\cite{Chachamis:2013rca,Caporale:2013bva,Chachamis:2012qw,Chachamis:2012fk,Chachamis:2011nz,Chachamis:2011rw}. 
The BFKL formalism is sensitive to collinear regions of phase space. The dominant double-log terms can be resummed
~\cite{Salam:1998tj}. 
%~\cite{Salam:1998tj,Ciafaloni:2003ek}. 
In~\cite{Vera:2005jt}, it was shown that this can be done  using 
\begin{eqnarray}
\theta \left(k_i^2-\lambda^2\right) \to \theta \left(k_i^2-\lambda^2\right)  + \sum_{n=1}^\infty 
\frac{\left(-\bar{\alpha}_s\right)^n}{2^n n! (n+1)!} \ln^{2n}{\left(\frac{\vec{k}_A^2}{\left(\vec{k}_A+\vec{k}_i\right)^2}\right)}. 
\label{SumBessel}
\end{eqnarray}
This resummation shows agreement with experimental results and good perturbative convergence
~\cite{Vera:2006un}. 
%~\cite{Vera:2006un,Vera:2007kn,Caporale:2007vs,Vera:2007dr,Caporale:2008fj,Hentschinski:2012kr,Hentschinski:2013id,Caporale:2013uva,Chachamis:2015ona}. 
In~\cite{Chachamis:2015zzp}
 we have implemented it in the {\tt BFKLex} Monte Carlo event generator and studied~\cite{Chachamis:2015ico} Mueller-Navelet~\cite{Mueller:1986ey} configurations. Three averages for the jets in each event were investigated: of the modulus of their transverse momentum, of their azimuthal angle and of the rapidity ratio between subsequent jets:
\begin{eqnarray}
\langle p_t \rangle ~=~ \frac{1}{N} \sum_{i=1}^{N} |k_i|;
\label{eq:observable1} \,\, \,\,\, \,
\langle \theta \rangle ~=~ \frac{1}{N} \sum_{i=1}^{N} \theta_i;
\label{eq:observable2} \, \, \,\,\, \,
\langle {\mathcal R}_y \rangle =~ \frac{1}{N+1}  \sum_{i=1}^{N+1} \frac{y_i}{y_{i-1}}.
\label{eq:observable3}
\end{eqnarray}
We studied the configurations with $k_a = 10$ GeV, $k_b = 20$ GeV and  $y_a-y_b = 4, 6, 8$.  
In  Fig.~\ref{Plots}
we show broad distributions in $\langle p_t \rangle$ with a maximal value at $\langle p_t \rangle \simeq 6$ GeV for $k_b=12$ and 8 GeV for $k_b=20$.  The cross section receives sizable contributions from  large transverse momentum jets.  We have taken a random value for the azimuthal angle between the two tagged jets which changes event by event. The $\langle \theta \rangle$ per event at which the remaining jets are produced is shown in Fig.~\ref{Plots} together with the mean ratios of rapidities  $\langle {\mathcal R}_y \rangle$.
The distributions peak at $\langle {\mathcal R}_y \rangle > 0.5$. Since they are broad, there are substantial contributions  from preasymptotic configurations. 
The observables here presented are important to establish if the pre-asymptotic effects are present in the data.  The advantage of the LHC to study this type of physics is the large available  energy together with high statistics, which allows for  strong kinematical cuts. 

\section{Infrared Effects in Cross Sections}

A pressing question in the BFKL formalism is to find the best treatment of the running of the coupling. A challenging proposal was put forward by Lipatov~\cite{lipatov86} and his collaborators where the kernel is modified in the infrared to allow for the existence of Regge poles instead of a cut in the complex angular momentum plane. To clarify the implications of this approach in~\cite{Ross:2016zwl} we have studied the diffusion for the typical transverse momentum in the BFKL ladder and its dependence on the choice of boundary conditions. 

\begin{figure}
\vspace{-.3cm}
\begin{center}
\includegraphics[width=7.0cm]{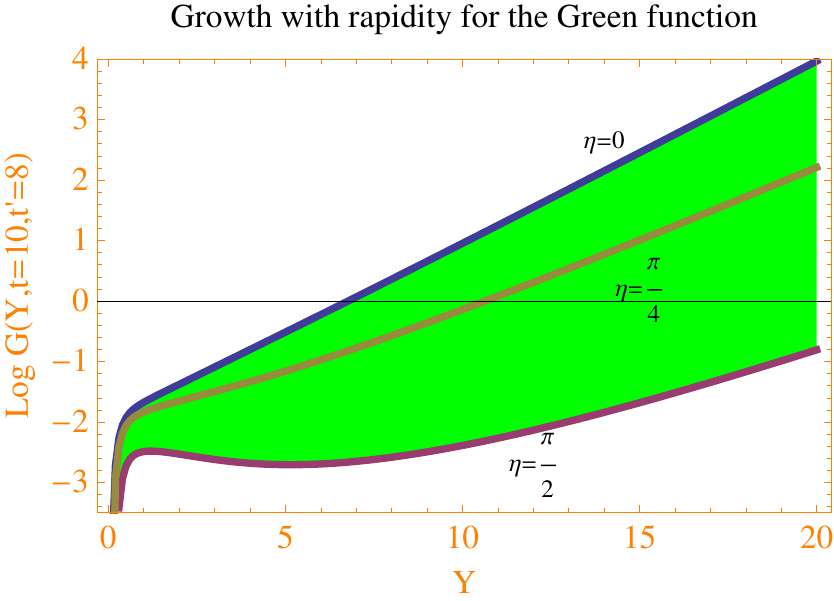}  \includegraphics[width=7.0cm]{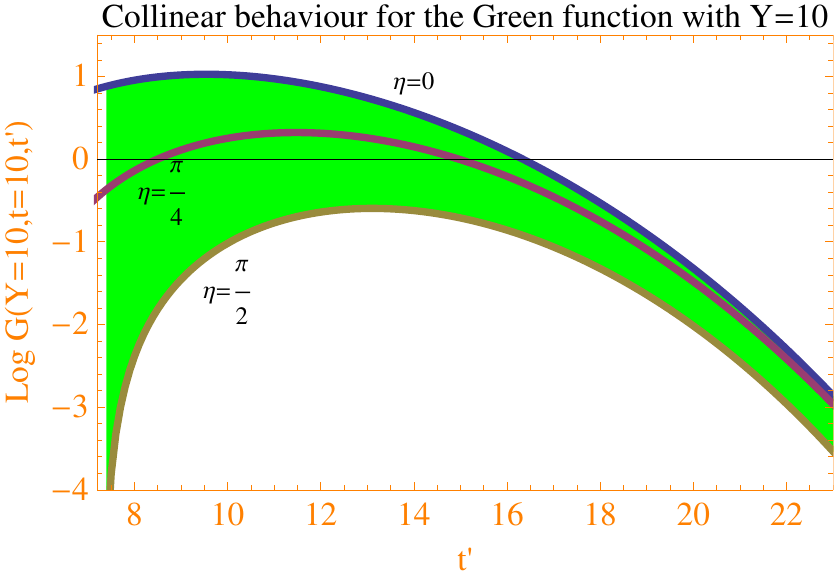}  
\end{center} 
\vspace{-.4cm}
\caption{Effect of non-perturbative phase on the gluon Green function.}
\label{GGFplot}
\end{figure}

\begin{figure}
\vspace{-.1cm}
\begin{center}
 \includegraphics[width=7.0cm]{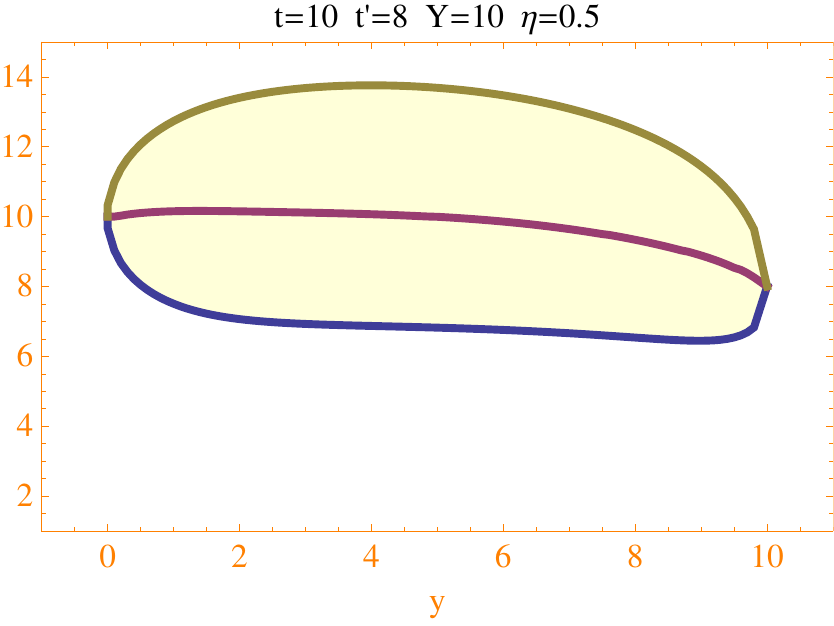} \\
 \includegraphics[width=7.0cm]{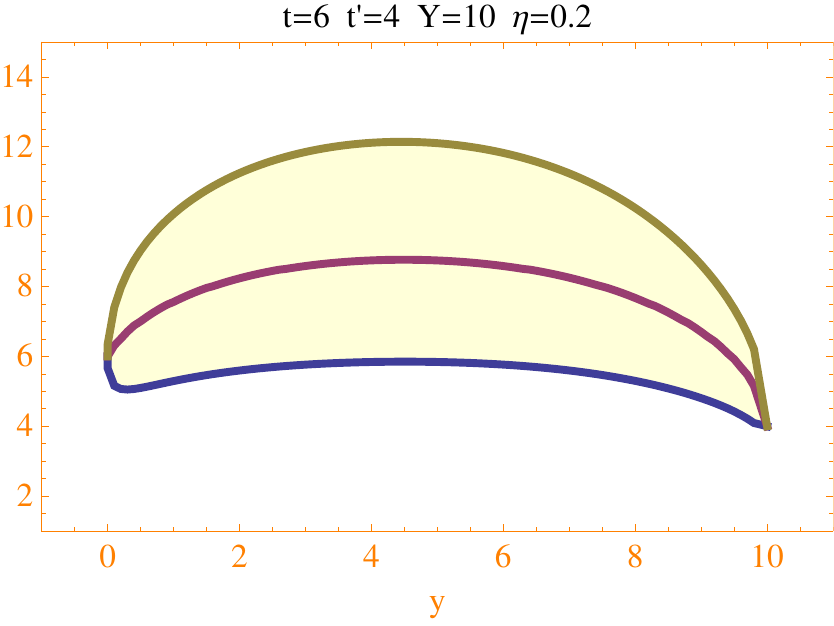}     
\end{center} 
\vspace{-.4cm}
\caption{Effect of the non-perturbative phase on the diffusion profile. }
\label{CigarsPositiveEta}
\end{figure}

Using the notation  $t_i \equiv \ln(k_i^2/\Lambda^2_{\mathrm{QCD}})$,  $ \bar{\alpha}_s \ \equiv \  \frac{C_A}{\pi}  \alpha_s$,  $ \bar{\alpha}_s(t) \ = \ \frac{1}{\bar{\beta_0} t} $, and a form of 
 running the coupling with hermitian kernel, the Green function equation is
\begin{equation} \frac{\partial}{\partial Y} {\cal G}(Y,t_1,t_2) 
\ = \ 
 \frac{1}{\sqrt{\bar{\beta_0} t_1}}
  \int dt \, {\cal K}(t_1,t)   \frac{1}{\sqrt{\bar{\beta_0} t}}    \, 
{\cal G}(Y,t,t_2) 
% \ = \ \delta\left(t_1-t_2\right) , \label{green_running}
\end{equation}
The Mellin transform of this Green function can be expressed in terms of 
 Airy functions~\cite{KLR1,KLR2},  
\begin{equation} {\cal G}_\omega\left( t_1,t_2\right) \ = \ 
 \frac{\pi}{4} 
\frac{\sqrt{t_1t_2}}{\omega^{1/3}} 
\left( \frac{\bar{\beta_0} }{14 \zeta(3)} \right)^{2/3} 
 Ai\left(z(t_1) \right) Bi\left(z(t_2) \right)  \theta\left(t_1-t_2\right)
 \, + \, t_1 \leftrightarrow t_2  \label{mellin-green-1}\end{equation}
with
 $$ z(t) \ \equiv \left(\frac{\bar{\beta_0} \omega}{14\zeta(3) }\right)^{1/3}
 \left(t-\frac{4\ln 2}{\bar{\beta_0} \omega} \right). $$
Still satisfying the requirement that the Green function vanish when $t_1 \to \infty$ or $t_2 \to \infty$ we can add to this function 
 any solution of the homogeneous equation with the same ultraviolet behaviour, {\it i.e.} we can replace the Airy function  $Bi(z)$ by
$$ \overline{Bi}(z) \ \equiv Bi(z) + c(\omega) Ai(z). $$ 
If we take $c(\omega)$ of the form
\begin{equation} c(\omega) \ = \ \cot\left(\eta- \frac{2}{3} \sqrt{\frac{\bar{\beta_0} \omega}{14\zeta(3)}} \left(
 \frac{4\ln 2}{\bar{\beta_0}\omega}-t_0\right)^{3/2}  \right),  
\label{phase} 
 \end{equation}
a set of discrete poles in  $\omega$ appear when
\begin{equation} 
\eta- \frac{2}{3} \sqrt{\frac{\bar{\beta_0} \omega}{14\zeta(3)}} \left(
 \frac{4\ln 2}{\bar{\beta_0}\omega}-t_0  \right)^{3/2}  \ =  - n \pi. 
 \label{NPphase}
 \end{equation}
This sets the phase at some infrared fixed point $t_0$  to $\frac{\pi}{4}+\eta$. The value of 
 $\eta$ is determined by the non-perturbative properties of QCD~\cite{lipatov86}. Using a quadratic approximation for the kernel, we have numerically inverted the Mellin transform and  found that the non-perturbative parameter $\eta$ affects the phase in Eq.~(\ref{NPphase}). Increasing the value of $\eta$ reduces the rise with energy of the gluon Green function, see top plot in Fig.~\ref{GGFplot}.
We study the collinear behaviour in RHS of the same figure. Note that the effect of $\eta$ is much more relevant at small values of $t$, which is expected, since the value of $\eta$ encodes the infrared behaviour.  This $t$-profile contains important information about the discrete pomeron approach since it allows for the study of  the diffusion in transverse scales as we show in Fig.~\ref{CigarsPositiveEta}.
When there are large external scales (bottom) we have a rather UV/IR symmetric diffusion. However we see a strong suppression of the diffusion towards the IR as we lower the external scales. Even the mean values of the distributions (central line) are pushed towards harder scales than the external ones.   All of this shows that  in the discrete pomeron approach there is an effective barrier screening the IR effects. We are not able to fix the non-perturbative phase from first principles. It will be very interesting to find the best fit to experimental observables using this free parameter as a new degree of freedom. The rich phenomenology in the multi-regge region available at the Large Hadron Collider~\cite{N.Cartiglia:2015gve} should help in the pursuit of this program. Higher order corrections can be implemented in this formalism as has been shown in~\cite{Ross:2016kzz} and~\cite{Kowalski:2017umu}.

\end{document}